\documentclass[twocolumn,showpacs,preprintnumbers,amsmath,amssymb]{revtex4}
\usepackage{graphicx}
\usepackage{dcolumn}
\usepackage{bm}
\usepackage{lipsum}
\usepackage{amsfonts}
\begin{document}

\title[]{Study of the $^7$Be($d$,$^3$He)$^6$Li* reaction at 5 MeV/u}

\author{Sk M. Ali$^1$}
\email{mustak@frib.msu.edu}
\altaffiliation {Present Address: FRIB, Michigan State University, East Lansing, MI 48824, USA}
\author{D. Gupta$^1$}%
\email{dhruba@jcbose.ac.in}
\author{K. Kundalia$^1$}
\author{S. Maity$^1$}
\author{Swapan K Saha$^1$}
\altaffiliation {Visiting faculty, School of Astrophysics, Presidency University, 86/1 College Street, Kolkata 700073, India}
\author{O. Tengblad$^2$}
\author{J.D.~Ovejas$^2$}
\author{A. Perea$^2$}
\author{I. Martel$^3$}		
\author{J. Cederkall$^4$}
\author{J. Park$^4$}
\altaffiliation{Present Address: Center for Exotic Nuclear Studies, Institute for Basic Science, 34126 Daejeon, South Korea}
\author{A.~M.~Moro$^5$}

\affiliation{$^1$Department of Physical Sciences, Bose Institute, EN 80, Sector V, Bidhannagar, Kolkata 700091, India}
\affiliation{$^2$Instituto de Estructura de la Materia $-$ CSIC, Serrano 113 bis, ES-28006 Madrid, Spain}
\affiliation{$^3$University of Huelva, Avenida Fuerzas Armadas sin numero Campus ``El Carmen", 21007, Huelva, Spain}
\affiliation{$^4$Department of Physics, Lund University, Box 118, SE-221 00 Lund, Sweden}

 \affiliation{$^5$Departamento de F\'{\i}sica At\'omica, Molecular y Nuclear, Facultad de F\'{\i}sica, Universidad de Sevilla, Apartado 1065, E-41080 Sevilla, Spain}

\begin{abstract}
The measurement of the $^7$Be($d$,$^3$He)$^6$Li* transfer cross section at 5 MeV/u is carried out. The population of the 2.186 MeV excited state of $^6$Li in this reaction channel is observed for the first time. The experimental angular distributions have been analyzed in the finite range DWBA and coupled-channel frameworks. The effect of the $^7$Be($d$,$^3$He)$^6$Li reaction on both the $^6$Li and $^7$Li abundances are investigated at the relevant big-bang nucleosynthesis energies. The excitation function is calculated by TALYS and normalized to the experimental data.  The $S$ factor of the ($d$,$^3$He) channel from the present work is about 50$\%$ lower than existing data 
at nearby energies. At big-bang energies, the $S$ factor is about three orders of magnitude smaller than that of the ($d,p$) channel. The ($d$,$^3$He) reaction rate is found to have a less than 0.1$\%$ effect on the $^{6,7}$Li abundances. 
    
\end{abstract}

\date{\today}

\maketitle

\section{Introduction}

In nuclear astrophysics, the cosmological lithium problem is widely known and the solution does not exist at present. The problem involves an anomaly in the observed $^7$Li abundance in metal-poor stars as compared to the big-bang nucleosynthesis (BBN) theory~\cite{CO04, SP82, BO10, FI11, CY16}. The BBN theory overestimates the $^7$Li abundance by a factor of $\sim$ 3 and substantial efforts have been devoted to solve this problem by studying the relevant nuclear reactions and searching for new resonances inside the Gamow window~\cite{AN05,HA13,BA16,DA18,LA19,RI19,AL22}. At present, it is believed that nuclear physics solutions to this problem are highly improbable~\cite{KI11,AL22}, and therefore solutions beyond standard model are contemplated~\cite{FI11,CY16}. 

A number of studies also point towards a second cosmological lithium problem involving $^6$Li. Here the BBN calculations give the $^6$Li/$^7$Li ratio $\sim$ 10$^{-5}$~\cite{SE04,MU16}, as compared to the observed values of $0.01-0.10$~\cite{MO17,GO19,WA22}. 
The $^6$Li abundance in metal-poor stars exhibit a plateau as a function of metallicity~\cite{NO97,AS06,AS08,AN14} similar to $^7$Li, suggesting a primordial origin. In contrast to $^7$Li, the BBN predictions here underestimate the observed abundance by a factor of $\sim$ 10$^3$~\cite{BO10,AN14}. However, more recent studies~\cite{MO17,GO19,WA22} have contradicted the $^6$Li detections in Spite plateau stars. These new analyses employ 3D non-local thermodynamic equilibrium (NLTE) modelling of the stellar atmosphere. 
The analyses provide upper limits to the isotopic ratio of $^6$Li/$^7$Li $<10\%$ at very low metallicity. The results are compatible with no detection of $^6$Li as well, and  thus supporting the negligible amount of $^6$Li predicted by BBN models. The results neither solve the cosmological $^7$Li problem nor worsen it and do not support exotic scenarios for significant $^6$Li production in the early Universe~\cite{KU08,JE09,LU21}. However, there are also a number of peculiar stars where the $^6$Li excess has been verified~\cite{TA17}.
It is to be noted that the determination of primordial $^6$Li abundance inferred from observations, requires high sensitivity spectrometers and improved stellar modelling. The 3D-NLTE analyses from different authors~\cite{TR17} do not yield the same $^6$Li/$^7$Li ratios and in some cases, there is also good agreement between 1D and 3D models.  

The disagreement in case of the lithium isotopes is of considerable importance, especially when one finds remarkable agreement between
BBN predictions and primordial abundances of $^2$H and $^4$He~\cite{FI11}. It may be noted that, simultaneous solutions to both the $^6$Li and $^7$Li problems have also been proposed through the existence of massive, negatively charged  particles binding to light nuclei in the early universe. The bound exotic particles would reduce the reaction Coulomb barriers, thereby extending BBN to lower temperatures~\cite{KU07}.  Thus $^{6,7}$Li problems may provide insight to important post-primordial scenarios.
 
In this context, a $^7$Be destruction reaction that may impact both the lithium problems simultaneously, is $^7$Be($d,^3$He)$^6$Li. This reaction not only produces $^6$Li but also destroys $^7$Be thereby decreasing $^7$Li abundance indirectly, as primordial $^7$Li mostly originated from $^7$Be. Sensitivity studies by Boyd~\textit{et al}.~\cite{BO10} concluded that the $^7$Be($d$,$^3$He)$^6$Li reaction could not change the BBN abundances of $^{6,7}$Li by more than 0.1\% assuming a factor of 1000 enhancement in the reaction rate. However, if the reaction were to proceed through a thermally populated first excited state, $^7$Be$^\ast$ with excitation energy of 429 keV, then it produced a 30\% decrease in the mass-7 abundance when the rates were artificially increased by thousand times. Broggini \textit{et al}.~\cite{BR12} also found that the effect of the $^7$Be($d$,$^3$He)$^6$Li reaction is too small to explain the observations of $^6$Li~\cite{AS06}. 
The study of the $^7$Be($d$,$^3$He)$^6$Li$^*$($\alpha$,d) reaction, with breakup of $^6$Li from its excited states, can also shed light on the
$^7$Be($d$,t)$^6$Be reaction and relevant resonances in $^6$Be~\cite{GU03,CH12}. At present, the only experimental data for $^7$Be($d$,$^3$He)$^6$Li reaction,
are by Li~\textit{et al}.~\cite{LI18}, at center of mass energies of $E_{cm}$= 4.0 and 6.7 MeV.

The authors in Ref.~\cite{LI18} initially considered the ($d$,$^3$He) reaction rate to be the same as the ($d,p$) rate. They also artificially multiplied the rate by a factor of 100. As a result, their BBN calculations show a 45\%  decrease in abundance of $^7$Li and 47\% increase in abundance of $^6$Li. The experiment by Li \textit{et al}.~\cite{LI18} used low $^7$Be beam intensities of 5000 pps. The reaction products $^3$He and $^6$Li were selected from $\Delta E-E$ spectrum using gates from Monte Carlo simulations of the  $^7$Be($d$,$^3$He)$^6$Li reaction. Although background subtraction was done using a carbon target, no kinematical signatures were shown to verify that the selected $^3$He and $^6$Li are indeed from the above reaction. Thus, experimental investigation of the $^7$Be($d$,$^3$He)$^6$Li reaction in further details is required. This would also help to ascertain the relative importance of the ($d,p$) and ($d,^3$He) channels affecting the lithium abundances. The ($d,p$) reaction cross sections at $E_{cm}= 7.8$ MeV have been reported in Ref.~\cite{AL22}, while the present paper involves the ($d,^3$He) reaction. 

\begin{figure}[h!]
    \centering 
    \includegraphics[width =0.5\textwidth]{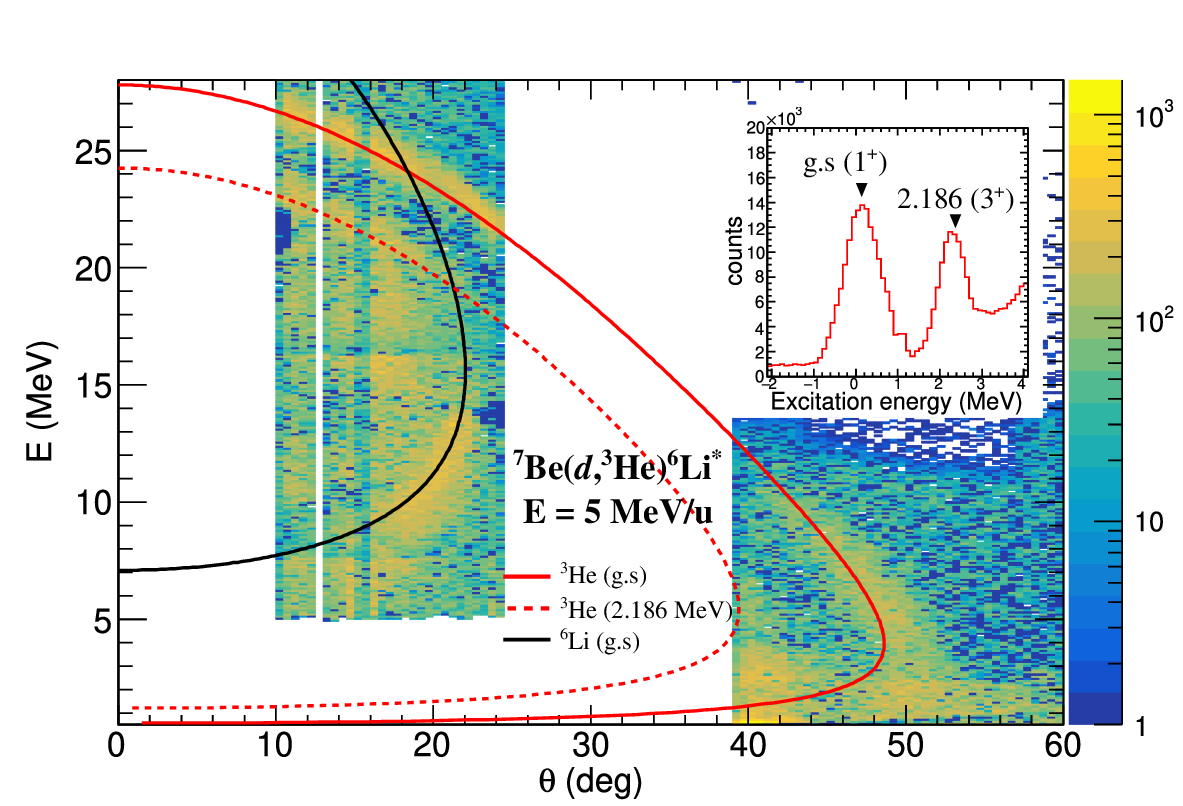}
    \caption{ $E$ vs $\theta$ plot of events from $^7$Be($d$,$^3$He)$^6$Li* reaction at 5 MeV/u. The kinematic lines corresponding to the $^3$He and $^6$Li bands at S3 covering 8$^\circ$-25$^\circ$ and W1 detectors covering above 40$^\circ$ are also shown (see text for details). The inset shows the excitation energy spectrum of $^6$Li.}
    \label{fig : E_Theta_3He}
\end{figure}

\section{Experiment}

The experiment was carried out at the HIE-ISOLDE radioactive ion beam facility of CERN. A 5 MeV/u $^7$Be beam of intensity $\sim$ 5 $\times$ 10$^5$ pps and energy resolution $\sim$ 168 keV was incident on a CD$_2$ target of thickness 15 $\mu$m. The details of the beam production and the experiment are described in~\cite{AL22}. 
The charged particles emitted from the reaction were detected by a compact array of silicon strip detectors. 
The forward angles from 8$^{\circ}$ $-$ 25$^{\circ}$ were covered by a 1000 $\mu$m thick annular detector (S3) with 24 rings and 32 spokes~\cite{AL22,MI}. A set of five detector telescopes assembled in a pentagon geometry covered the angular range 40$^{\circ}$ - 80$^{\circ}$. The thickness of the thin $\Delta E$ (W1) and thick $E$(MSX25) layers were 60 $\mu$m and 1500 $\mu$m respectively. The W1 detector is a 16x16 silicon strip detector while the MSX25 is an unsegmented silicon pad detector~\cite{AL22,MI}. Two more telescopes covered the back angles from 127$^{\circ}$ - 165$^{\circ}$~\cite{AL22}, but are not relevant for the present work. The size of the detector strips and the detector to target distance imposed an angular resolution of $\sim 1^\circ$ in S3 and $2^\circ$ in W1. The energy calibrations of the detectors were carried out by a mixed $\alpha$-source as well as by elastic peaks from the Rutherford scattering of ${5}$ MeV/u $^{7}$Be and ${5.15}$ MeV/u $^{12}$C beams on a $^{208}$Pb target. A 15 $\mu$m thick CH$_2$ target and a 1 mg/cm$^2$ $^{208}$Pb target were used for background measurements and normalization respectively. To normalize the present data, one of the forward rings of S3 was used as a monitor detector, and the $^7$Be + $^{208}$Pb elastic scattering data from this ring was assumed to be the Rutherford cross section. In addition to statistical uncertainty, the present data also include a 10$\%$ error due to target thickness and a 10$\%$ error due to beam intensity~\cite{KU22}.
 
\section{Results}

At the forward angles covered by S3, the $^7$Be($d$,$^3$He)$^6$Li reaction can give rise to events where both $^3$He and $^6$Li are incident on S3. These events are identified by imposing the condition of two hits at S3 and no hit at the W1 pentagon detectors. In addition to CH$_2$ background subtraction, the experimental data were further filtered by applying a gate on the total energy of the two hits. The total energy was obtained from Monte Carlo simulations of the reaction. The simulations, using the package NPTool~\cite{MA16}, took into account the beam energy spread, geometry and resolution of the detectors as well as energy and angular straggling in the target and detectors. The resultant energy ($E$) vs scattering angle ($\theta$) plot in Fig.~\ref{fig : E_Theta_3He} shows a clear band of $^3$He from $^7$Be($d$,$^3$He)$^6$Li reaction. The $^6$Li detected in coincidence with $^3$He can also be seen in the figure. 
The kinematics for $^3$He and $^6$Li are shown in Fig.~\ref{fig : E_Theta_3He} by the red and black lines respectively. Another $^3$He band is also seen corresponding to the 2.186 MeV excitation of $^6$Li. The population of this state is observed for the first time in the $^7$Be($d$,$^3$He)$^6$Li$^\ast$ reaction channel. The 2.186 MeV state being above the $^6$Li breakup threshold of 1.474 MeV, breaks up into $\alpha$ and $d$. This $^3$He band is seen with the red-dotted kinematic line in Fig.~\ref{fig : E_Theta_3He}. It becomes more distinct when we consider the triple-coincidence ($^3$He-$\alpha$-$d$), identifying the $^3$He in coincidence with two hits at S3. 

The $^7$Be($d$,$^3$He)$^6$Li reaction can also give rise to events where the $^3$He is incident on W1 while the corresponding $^6$Li falls on S3. For such events, multiplicity condition of one hit at S3 and one hit at W1  were considered. The $^3$He nuclei corresponding to the ground state (g.s) of $^6$Li are completely stopped at $\Delta E$ (W1) and are detected up to an angle $\sim$ $50^\circ$, 
as can be seen from Fig.~\ref{fig : E_Theta_3He}. The $^6$Li corresponding to these $^3$He were also detected and analyzed. However, for the sake of clarity, these $^6$Li are not shown in Fig.~\ref{fig : E_Theta_3He}. For the 2.186 MeV state, the $^3$He are barely detected at W1, as can be seen by the red-dotted line around $40^\circ$ in Fig.~\ref{fig : E_Theta_3He}.  These events are therefore not considered in the analysis. Fig.~\ref{fig : E_Theta_3He} thus shows the superimposed contribution of three types of events that are analysed separately. These are events for (1) two S3 hits with no W1 hit, (2) three S3 hits and (3) one S3 and one W1 hits. The relevant efficiency corrections for cross section calculations are also separately done as explained later. The $^3$He may also originate from other channels like breakup of $^7$Be. These $^3$He traverse the $\Delta E$ detectors, stopping at the $E$ detectors. It may be noted that, these events were vetoed in the present analysis. The inset in Fig.~\ref{fig : E_Theta_3He} shows the excitation energy spectrum of $^6$Li. If we further gate on the $^3$He bands, the background in the excitation energy spectrum goes away.
\begin{figure}[h!]
    \centering 
    \includegraphics[width =0.5\textwidth]{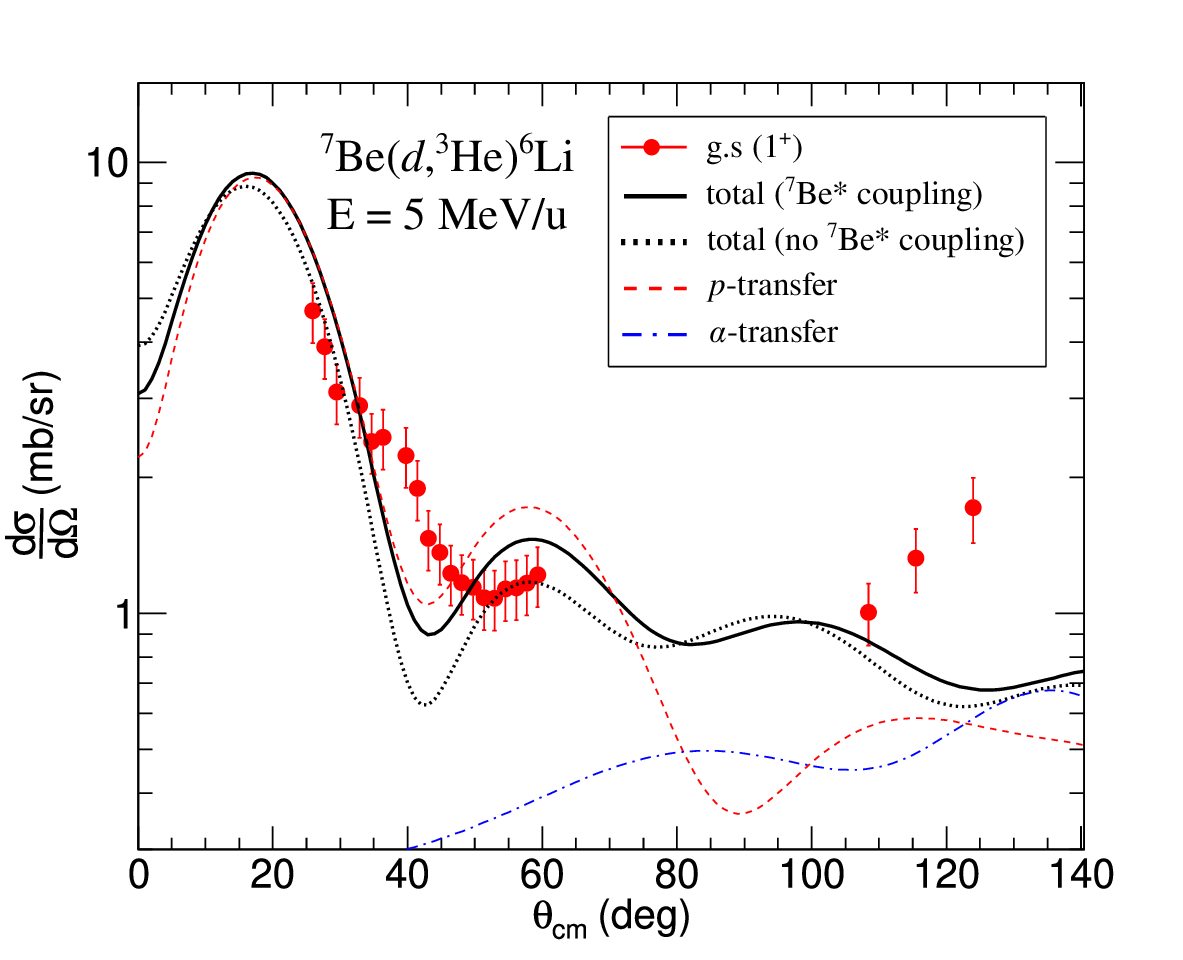}
    \caption{Angular distribution for the ground state of $^6$Li from $^7$Be($d$,$^3$He)$^6$Li reaction at 5 MeV/u. The CCBA calculations with (without) the coupling of the $^7$Be$^\ast$(1/2$^-$) excitation at 0.429 keV are shown by the black solid (dotted) lines. The separate $p$ and $\alpha$-transfer contributions with the coupling are also shown by the red dotted and blue dash-dotted lines respectively.}
    \label{fig : xs}
\end{figure}

\begin{figure}[h!]
    \centering 
    \includegraphics[width =0.5\textwidth]{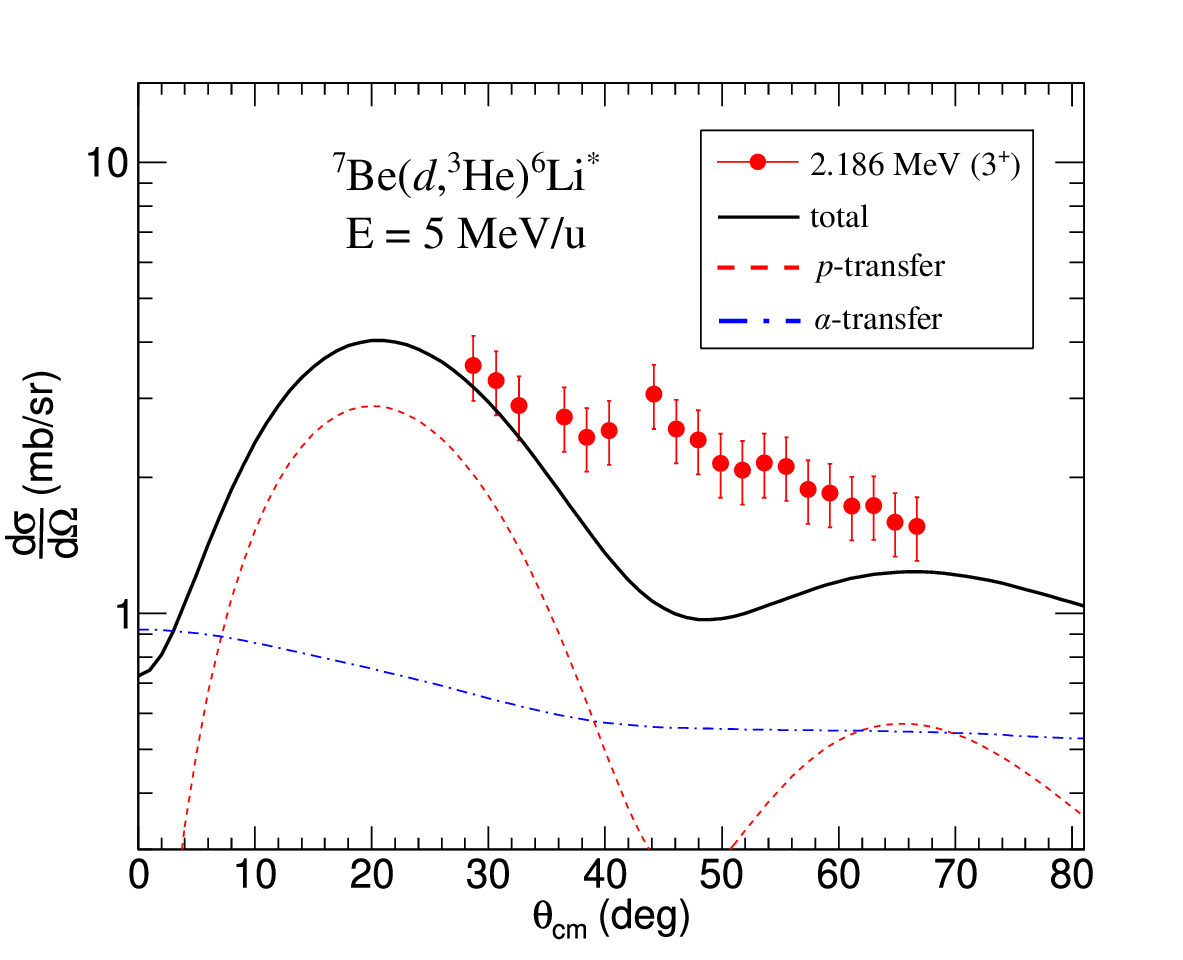}
    \caption{Angular distribution for the 2.186 MeV excited state of $^6$Li from  $^7$Be($d$,$^3$He)$^6$Li$^\ast$ reaction at 5 MeV/u. The CCBA calculation is shown by the black solid line. The separate $p$ and $\alpha$-transfer contributions are shown by the red dotted and blue dash-dotted lines respectively.}
    \label{fig : Ex_xs}
\end{figure}

The $^7$Be($d,^3$He)$^6$Li* angular distributions for the g.s ($1^+$) and 2.186 MeV($3^+$) state of $^6$Li are obtained from $^3$He counts in S3 and W1 detectors. For the g.s, there is no loss of $^3$He counts as all the $^3$He and $^6$Li are detected in coincidence at S3. However, for $^6$Li($3^+$), the $^3$He counts are corrected by taking into account the coincidence efficiency of simultaneous detection of $^3$He with $\alpha$ and $d$ at S3 from NPTool simulations. The absolute normalization for the cross sections were obtained from Rutherford scattering with a $^{208}$Pb target~\cite{AL22}. The angular distributions for the g.s and 2.186 MeV state of $^6$Li are shown in Figs.~\ref{fig : xs}~and~\ref{fig : Ex_xs} respectively. Initially, finite range distorted-wave Born approximation (DWBA) calculations were carried out using FRESCO~\cite{TH88}, assuming one-step $p-$transfer ($d,^3$He) and $\alpha$-transfer ($d,^6$Li). For the $p-$transfer, a $p$ + $^6$Li configuration of $^7$Be is considered, with the proton occupying a $1p_{3/2}$ or $1p_{1/2}$ configuration. The proton transferred to $d$ is in 1$s_{1/2}$, forming $^3$He. The optical model potential (OMP) parameters for $d$ + $^7$Be entrance channel are obtained in the same experiment by fitting the elastic scattering data~\cite{AL22}. The $^3$He + $^6$Li exit channel potential~\cite{LU68}, $d$ + $^6$Li core-core potential~\cite{BI65}, and $p$ + $^6$Li~\cite{BA80} and $d+p$~\cite{SA20} binding potentials, used in the present work are listed in Table~\ref{tab:OMP}. To investigate the possible influence of the $^7$Be first excited state on the calculated transfer cross sections, we have carried out coupled-channel calculations (CCBA)~\cite{SA83} including the $^7$Be$^\ast$(1/2$^-$) excitation at 0.429 keV. These calculations are explained later and shown in Figs.~\ref{fig : xs}~and~\ref{fig : Ex_xs}.

The proton spectroscopic factors used in the calculations are 0.438 and 0.225 for $p_{3/2}$ and $p_{1/2}$ components in $^7$Be = $p$ + $^6$Li($1^+)$ respectively~\cite{BR11} as given in Table~\ref{tab:SF}. For the $^7$Be = $p$ + $^6$Li($3^+)$ configuration, the proton spectroscopic factor is 0.457~\cite{BR11}. The spectroscopic factor for $^3$He in $d+p$ configuration is taken as 1.16~\cite{KI20}. 
\noindent For $\alpha-$transfer, the $^7$Be is assumed to have the $\alpha$ + $^3$He configuration in a $2P$ state of relative motion. The transferred $\alpha$ forms $^6$Li with $\alpha + d$ configuration in a $2S$ state. The spectroscopic factor for $^7$Be in $\alpha$ + $^3$He configuration is taken as 1.19~\cite{RU96}. 
According to the three-body calculations of Ref.~\cite{WA15}, the probability for the $^6$Li(gs) = $\alpha + d$  and $^6$Li($3^+$) = $\alpha + d$ configurations are 0.70 and 0.72 respectively. In the CCBA calculations presented here, we adopted these values for the corresponding spectroscopic factors. This should be a good approximation as long as antisymmetrization effects, not included in three-body calculations, are small. The core-core $d$ + $^3$He potential~\cite{PE76}, and binding potentials for $\alpha$ + $^3$He~\cite{BU88} and $\alpha$ + $d$ systems~\cite{FE15} are also given in Table~\ref{tab:OMP}. It may be noted that the $^6$Li($3^+$) state is populated by both $p$-transfer and $\alpha$-transfer. However, $^6$Li($3^+$) is unbound, being above the breakup threshold. To include its contribution in the $\alpha$-transfer calculations, a continuum bin is defined to be centred at the resonance energy of 0.71 MeV above the continuum. The width of the bin is taken as 0.7 MeV, as the experimental energy resolution is $\sim$ 660 keV. The $p$ and $\alpha$-transfer contributions are added coherently to give the total cross section. 

\begin{figure}[h!]
    \centering 
    \includegraphics[width =0.5\textwidth]{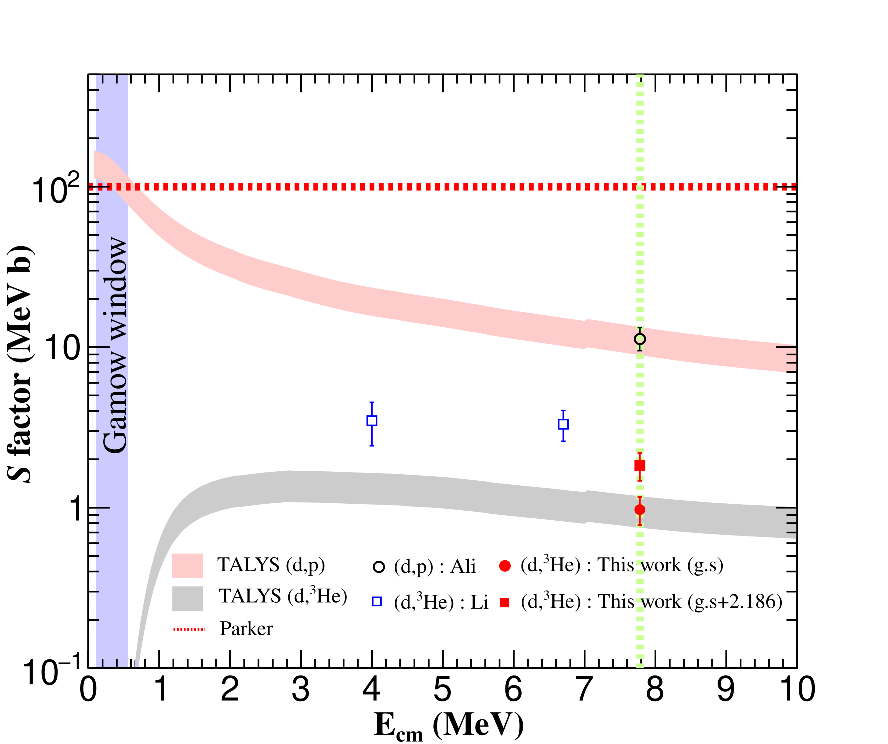}
    \caption{Astrophysical $S$ factor for the $^7$Be($d$,$^3$He)$^6$Li reaction. The red solid circle and square correspond to the present work while the blue open squares correspond to the data from Li \textit{et al}.~\cite{LI18}. The open black circle corresponds to ($d$,$p$) data~\cite{AL22}. The red and gray bands are TALYS calculations for ($d$,$p$) and ($d$,$^3$He) channels respectively, normalized to the present data at 7.8 MeV (green vertical line). The red dotted line is the ($d$,$p$) estimate by Parker~\cite{PA72}.}
    \label{fig : sfactor}
\end{figure}

In the CCBA calculations~\cite{SA83}, a collective model has been used to couple the 3/2$^-$ ground state and 1/2$^-$ excited state of $^7$Be. The Coulomb and nuclear potentials are deformed assuming the rotor model for $^7$Be. For the Coulomb part, the intrinsic reduced matrix element is taken as 11.4 $e$fm$^2$, as deduced from the $B(E2;3/2^- \rightarrow 1/2^-)$ value quoted in Ref.~\cite{HE19}. Assuming a pure rotor model, and equal mass and charge deformations, this leads to the nuclear deformation length $\delta_2$ = 2 fm. The proton spectroscopic factors for the $^7$Be$^\ast$(1/2$^-$) state are taken as 0.854 and 0.038 for 1$p_{3/2}$ and 1$p_{1/2}$ configurations respectively. They are obtained from the Shell model calculations with Cohen-Kurath interaction~\cite{CO65}. The bare OMPs for the $d$ + $^7$Be system~\cite{AL22} are re-adjusted due to the inclusion of deformation couplings in the entrance channel. The code FRESCO is used to adjust the depths and radii of the volume real and surface imaginary parts to best fit the elastic angular distribution. The modified OMPs are also included in Table~\ref{tab:OMP}. The CCBA calculation for the ground state with (without) $^7$Be$^\ast$(1/2$^-$) coupling is shown by the black solid (dotted) line in Fig.~\ref{fig : xs}. It is observed that the addition of the $^7$Be$^\ast$(1/2$^-$) state does not alter much the CCBA result, reducing only the dip of the first minimum. For the $^7$Be($d$,$^3$He)$^6$Li$^\ast$ calculation, the direct coupling between the $^7$Be excited state and the $^6$Li(3$^+$) final state is not possible within the restricted $p$-shell orbital considered in the Cohen-Kurath modelspace. Thus, the CCBA calculation for $^7$Be($d$,$^3$He)$^6$Li$^\ast$ reaction does not include the $^7$Be$^\ast$(1/2$^-$) coupling, and is shown by the black solid line in Fig.~\ref{fig : Ex_xs}. The contributions from $p$ and $\alpha$-transfer are also shown separately by the red dotted and blue dash-dotted lines respectively. The apparent disagreement between the calculated and measured angular distributions for the population of the $^6$Li($3^+)$ state might be due to the ambiguities of the optical potentials involved in the CCBA calculations or to the effect of channels not included in the calculations, such as the coupling to higher states in $^7$Be or between the $^6$Li states.

The total cross section $\sigma_\textrm{exp}$ leading to the g.s and 2.186 MeV state of $^6$Li are given in Table~\ref{tab:CS}. They are obtained by integrating the angular distributions in Figs.~\ref{fig : xs}~and~\ref{fig : Ex_xs} by using LISE++~\cite{TA16}. The error in the total cross section ($\Delta \sigma$) is obtained from the variation in total cross section considering the upper error limit $(d\sigma/d\Omega)_\textrm{high}$ and lower error limit  $(d\sigma/d\Omega)_\textrm{low}$ of the angular distribution data.  The error is given as $\Delta \sigma = (\sigma_\textrm{high} -\sigma_\textrm{low})/2$, where
$\sigma_\textrm{high/low} = 2\pi \int_{0}^{\pi}(d\sigma/d\Omega)_\textrm{high/low} (\sin\theta d\theta)$. The integrated total cross sections from CCBA calculations $\sigma_\textrm{CCBA}$ are also included in the table for comparison. The reason for the variation in the data and calculations for the 2.186 MeV state, is probably the absence of data at $\theta_{cm}>$ 70$^\circ$. To estimate the reaction cross section at the Gamow window ($T=0.5-1$ GK, $E_{cm}=0.11-0.56$ MeV), the excitation function was calculated with TALYS 1.95~\cite{KO19}. The calculations are then normalized to the present data at $E_{cm}=7.8$ MeV with the experimental uncertainty limits, shown by the coloured bands (Fig.~\ref{fig : sfactor}). The $Q$ value of this reaction is $-0.11$ MeV as compared to the high $Q$ value of +16.67 MeV in case of $^7$Be($d,p$)$^8$Be(2$\alpha$). For the $^7$Be($d$,$^3$He)$^6$Li$^\ast$ reaction to the 2.186 MeV state, the TALYS calculations give zero cross section below $E_{cm}= 0.71$ MeV. The energies in the Gamow window are below the reaction threshold to populate the 2.186 MeV state of $^6$Li. So this state is not considered in the $S$ factor normalization. The gray band in Fig.~\ref{fig : sfactor} shows the $S$ factor for the ($d$,$^3$He) channel from TALYS calculations with experimental uncertainties. The earlier work by Li \textit{et al}.~\cite{LI18} did not give any estimate of systematic uncertainties associated with the extrapolation of the cross sections to lower energies. The input parameters included in the TALYS calculations are the optical model potential parameters (OMP) and different level density models.  The effect of OMP is less than 10$\%$ on the $S$ factor. However, the choice of different models for level densities in TALYS can significantly impact the value of the $S$ factor inside the Gamow window. The normalized $S$ factor inside the Gamow window can vary from $0.001 - 0.04$ MeV b due to different level density models.
 
In our earlier work, the ($d,p$) measurements~\cite{AL22} including the contribution of the 16.63 MeV state led to a maximum total $S$ factor $\sim$ 167 MeV b. However, this has less than 1$\%$ effect in reducing the $^7$Li abundance from BBN calculations. From Fig.~\ref{fig : sfactor}, we see that the ($d$,$^3$He) $S$ factor extrapolated to BBN energies is more than three orders of magnitude lower than the ($d,p$) $S$ factor. Thus, it can be inferred that the contribution of the ($d$,$^3$He) channel is negligible as compared to the ($d,p$) channel in affecting the lithium abundance anomaly.
This agrees very well with the calculations in~\cite{BO10}. We also find that the total $S$ factor from the present measurements is almost 50$\%$ lower than earlier measurements~\cite{LI18} at nearby energies. The reason for this discrepancy may be lower statistics and contribution of other channels in the work of Li \textit{et al}~\cite{LI18}.\\
 
\begin{figure}[h!]
    \centering 
    \includegraphics[width =0.5\textwidth]{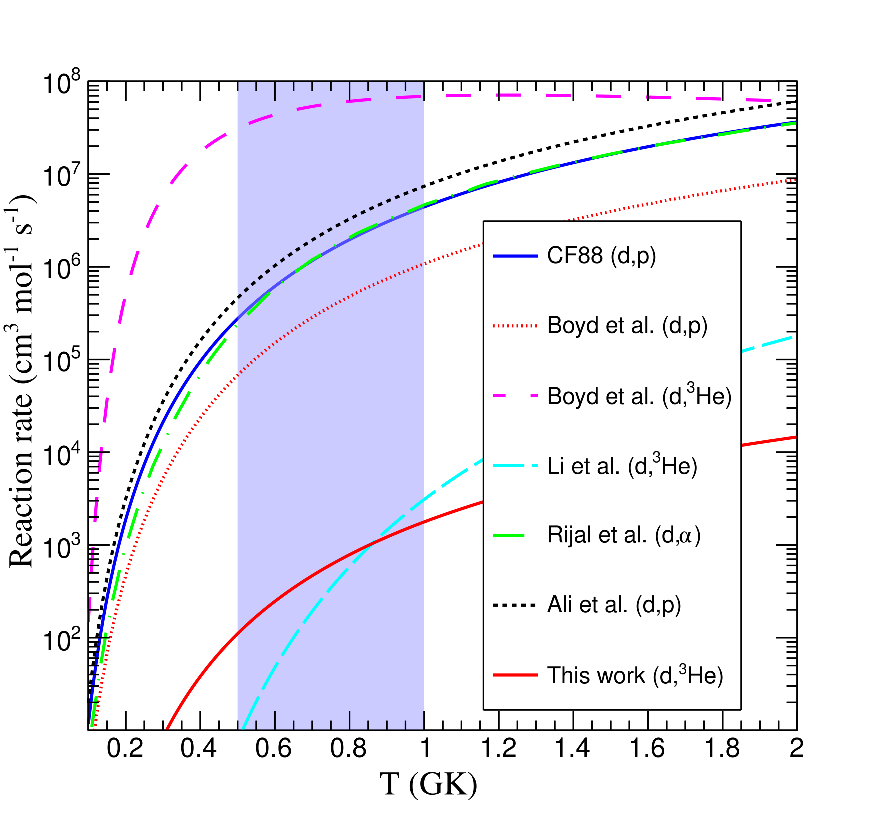}
    \caption{ The $^7$Be($d$,$^3$He)$^6$Li reaction rate from the present work (solid red line) as a function of temperature. For comparison, the rates from previous experimental and theoretical works are also shown. The $^7$Be($d,p$)$^8$Be$^\ast$ rate from ~\cite{AL22} is shown with black dotted lines. The blue band indicates the temperature range of interest for BBN ($0.5\leq T\leq 1$ GK).}
    \label{fig : reaction rate}
\end{figure}

Now, to study the impact of the above cross sections on the lithium production, we calculate the astrophysical reaction rate $N_A<\sigma~v>$~\cite{IL07}, where
\begin{align}
<\sigma~v>~=
      \bigg(\frac{8}{\pi \mu} \bigg)^{1/2} \frac{1}{(kT)^{3/2}} \int_{0}^{\infty} \sigma(E) E e^{-E/kT} dE.
\end{align}
Here, $N_A$ is the Avogadro number, $\sigma$ is the total cross section, $v$ is the velocity, 
$\sigma(E)$ is the excitation function, $\mu$ is the reduced mass of the $^7$Be + d system and $k$ is the Boltzmann constant. In Fig.~\ref{fig : reaction rate}, the reaction rates of ($d,p$) and ($d$,$^3$He) considering the maximum $S$ factor of 167 MeV b~\cite{AL22} and 0.04 MeV b (present work) at E$_{cm}$ = 0.22 MeV (Gamow peak at T = 0.5 GK), are compared with previous experimental and theoretical works. Although the total $S$ factor from the present work is lower as compared to that in~\cite{LI18}, the reaction rate at BBN energies is greater since we have considered the reaction rate with the maximum possible $S$ factor taking into account all the level density models in TALYS. The ($d$,$^3$He) reaction rate from Boyd \textit{et al}.~\cite{BO10}, shown in Fig.~\ref{fig : reaction rate}, corresponds to $^7$Be$^\ast$($d$,$^3$He)$^6$Li reaction proceeding through the thermally populated first excited state of $^7$Be as mentioned in the introduction.\\
To have an estimate of the abundance ratios of the lithium isotopes, the code \texttt{PArthENoPE} 3.0~\cite{GA22} was used to carry out the standard BBN calculations. It can take the input of reaction networks of up to 100 reactions. The code was run with the following conditions : the neutron average lifetime $\tau_n=879.4$ s~\cite{ZY20}, $N_\nu=3$ neutrino species~\cite{ZY20} and the baryon-to-photon ratio $\eta=6.10 \times 10^{-10}$~\cite{AL22}. 
If we consider the ($d,p$) reaction rate from CF88~\cite{CF88} and do not incorporate ($d$,$^3$He) reaction, the BBN calculations provide $^6$Li/H $=$ $1.8 \times 10^{-14}$ at BBN energies. Incorporating the ($d,p$) reaction rate from Ali {\it \textit{et al}.}~\cite{AL22} and ($d$,$^3$He) rate from the present work into the code, the $^6$Li and $^7$Li abundances change respectively by $\sim 0.01\%$ and $\sim 0.1\%$. 
The $^6$Li/$^7$Li abundance ratio from the BBN calculations is $\sim 10^{-4}$ which is well below the observed ratio of $0.01-0.10$~\cite{MO17,GO19,WA22}. A thousand fold increase of both the ($d,p$) and ($d$,$^3$He) reaction rates from our work can give rise to 35$\%$ decrease in the abundance of $^7$Li and only a 0.9$\%$ increase in abundance of $^6$Li. So it can be concluded that the effect of the $^7$Be($d$,$^3$He)$^6$Li reaction to any of the lithium problems is insignificant.
\begin{table*}[th!]
\caption{\label{tab:OMP}Potential parameters used in the present work. $V$ and $W$ are the real and imaginary depths in MeV, $r$ and $a$ are the radius and diffuseness in fm. $R_x = r_x A^{1/3}$ fm ($x=V, W, S, SO, C$).}
\begin{ruledtabular}
\begin{tabular}{c  c  c c c c   c  c  c  c  c  c  c  c  c}
Channel &$V$ &$r_{V}$ &$a_{V}$ &$W_{V}$ &$r_{W}$ &$a_{W}$ &$W_{S}$ &$r_{S}$ &$a_{S}$ &$V_{SO}$ &$r_{SO}$ &$a_{SO}$ &$r_C$  &Ref.\\
\hline 
$d$ + $^7$Be &80.98 &1.35   &0.83 &$-$ &$-$ &$-$    &36.91  &2.21   &0.10  &2.08    &0.49  &0.42 &1.30     &\cite{AL22}\\
             &79.47 &1.32 &0.83 &$-$ &$-$ &$-$      &31.46  &2.37   &0.10  &2.08    &0.49  &0.42 &1.30     &This work\\ 
$^3$He + $^6$Li &140.0  &1.20   &0.75 &30.0 &2.30 &0.50    &$-$  &$-$   &$-$  &$-$  &$-$  &$-$ &1.30   &\cite{LU68}\\
$d$ + $^6$Li   &78.0  &1.04   &0.95 &30.0 &0.85 &0.85   &$-$  &$-$    &$-$    &12.50  &1.04    &0.95  &1.04  &\cite{BI65}\\
$d$ + $^{3}$He &80.2 &2.36 &0.50 &3.7 &2.36 &0.50  &$-$  &$-$   &$-$  &$-$  &$-$  &$-$ &1.30 &\cite{PE76}\\
$\alpha$ + $d$ (g.s) &77.47 &1.05 &0.65 &$-$   &$-$  &$-$ &$-$   &$-$  &$-$ &$-$   &$-$  &$-$ &1.05 &\cite{FE15}\\
$\alpha$ + $d$ (3$^+$) &85.39  &1.05 &0.65 &$-$   &$-$  &$-$ &$-$   &$-$  &$-$ &$-$   &$-$  &$-$ &1.05 &\cite{FE15}\\
$p$ + $^6$Li &\footnote{{Varied to match separation energy}} &1.25 &0.65  &$-$   &$-$  &$-$ &$-$   &$-$  &$-$ &$-$   &$-$  &$-$ &$-$ &\cite{BA80}\\
$d$ + $p$ &\footnotemark[1] &1.25 &0.65 &$-$   &$-$  &$-$ &$-$   &$-$  &$-$ &$-$   &$-$  &$-$ &$-$ &\cite{SA20}\\
$\alpha$ + $^3$He\footnote{$V(r)=-(V+4\alpha V_{SO} \vec{l}\cdot \vec{s})\exp{(-\alpha r^2)}, \alpha=0.15747$ fm$^{-2}$} &83.78 &$-$ &$-$ &$-$ &$-$ &$-$&$-$ &$-$&$-$&$1.003$&$-$&$-$&3.095 &\cite{BU88}\\
\end{tabular}
\end{ruledtabular}
\end{table*}
\begin{table}[h!]
\caption{\label{tab:SF}Spectroscopic factors ($C^2S$) for $A=B+x$ systems used in the CCBA calculations of the present work. }
\begin{ruledtabular}
\begin{tabular}{c c c c c c } 
$A$ &$B$ &$x$    &{$nlj$}  &{$C^2S$}  &Ref. \\ 
\hline                    
$^7$Be &$^6$Li &$p$  &$1p_{3/2}$  &$0.438$ &\cite{BR11}\\
       &       &     &$1p_{1/2}$  &$0.225$ &\cite{BR11}\\
$^7$Be &$^6$Li$^\ast$ &$p$  &$1p_{3/2}$  &$0.457$ &\cite{BR11}\\
$^3$He &$d$  &$p$  &$1s_{1/2}$  &$1.16$ &\cite{KI20}\\
$^7$Be &$\alpha$  &$^3$He &$2P_{1}$ &1.19 &\cite{RU96}\\
$^6$Li (g.s) &$\alpha$ &$d$ &$2S_{1}$ &0.70 &\cite{WA15}\\
$^6$Li (3$^+$) &$\alpha$ &$d$ &$2S_{1}$ &0.72 &\cite{WA15}\\
\end{tabular}
\end{ruledtabular}
\end{table}
\begin{table}[h!]
\caption{\label{tab:CS}The total cross section obtained from CCBA calculations $(\sigma_\textrm{CCBA})$ and the experimental data $(\sigma_\textrm{exp})$ for $^7$Be($d$,$^3$He)$^6$Li*.}
\begin{ruledtabular}
\begin{tabular}{c c c c } 
E$_x$ (MeV)  &{$J^\pi$}  &{$\sigma_\textrm{CCBA}$ (mb)}  &{$\sigma_\textrm{exp}$ (mb)}   \\ 
\hline                    
0.0    &$1^+$ &$17.83$    &$21.2 \pm 4.2$  \\

2.186  &$3^+$ &$22.40$    &$18.6 \pm 3.7$  \\
\end{tabular}
\end{ruledtabular}
\end{table}

The present work is a significant improvement on the only experimental work~\cite{LI18} involving  the $^7$Be($d$,$^3$He)$^6$Li reaction. We studied the $(d,p)$ and ($d$,$^3$He) reactions at the same center of mass energy, quantitatively describing their relative contribution in the context of the lithium abundance. Although sensitivity studies have shown that the reaction does not influence the lithium abundance problem~\cite{BO10}, they did not involve any experimental constraints. We also measured, for the first time, transfer reaction cross sections to the 2.186 MeV excited state of $^6$Li, breaking up into $\alpha$ and $d$. This may have important consequences in similar studies involving the resonance states of the unbound $^6$Be nucleus~\cite{GU03,CH12}. 
The angular distributions of the ground state and the excited state can be used to constrain the Asymptotic Normalization Coefficients (ANC) of $^7$Be$\rightarrow$$^3$He+$^4$He and $^7$Be$\rightarrow$$^6$Li+$p$ systems, useful in the study of astrophysically relevant reactions such as $^3$He($\alpha$,$\gamma$)$^7$Be and $^6$Li($p$,$\gamma$)$^7$Be~\cite{LI18b,KI20,KI21}.

\section{Conclusion}
We report the measurement of the $^7$Be($d$,$^3$He)$^6$Li* transfer cross section at 5 MeV/u, observing the 2.186 MeV excited state of $^6$Li for the first time in this reaction. The effect of the $^7$Be($d$,$^3$He)$^6$Li reaction on the lithium abundances was investigated at the relevant BBN energies. The excitation function was determined by normalizing the TALYS calculations to the experimental total cross section of 21.2 $\pm$ 4.2 mb (present work) for the ground state. The $S$ factor of the ($d$,$^3$He) channel is found to be $\sim$ 50$\%$ lower than extant data at nearby energies of 4.0 and 6.7 MeV. It is also smaller by three orders of magnitude than ($d,p$)~\cite{AL22} at BBN energies. The ($d$,$^3$He) channel is found to have a negligible influence ($\leq$ 0.1 $\%$) on the lithium abundances. The present work provides a substantially improved cross section measurement of the  $^7$Be($d$,$^3$He)$^6$Li* reaction, important for similar studies in $^6$Be as well as constraining ANC values in astrophysically relevant reactions.\\

\noindent
{\large{\bf Acknowledgement}}\\

\noindent
The authors thank the ISOLDE engineers in charge, RILIS team and Target Group at CERN for their support. D. Gupta acknowledges research funding from the European Union's Horizon 2020 research and innovation programme under grant agreement no. 654002 (ENSAR2) and ISRO, Government of India under grant no.
ISRO/RES/2/378/15$-$16. O. Tengblad would like to acknowledge the support by the Spanish Funding Agency (AEI / FEDER, EU) under the project PID2019-104390GB-I00. 
I. Martel would like to acknowledge the support by the Ministry of Science, Innovation and Universities of Spain (Grant No. PGC2018-095640-B-I00). J. Cederkall acknowledges grants from the Swedish Research Council (VR) under contract numbers VR-2017-00637  and  VR-2017-03986 as well as grants from the Royal Physiographical Society. J. Park would like to acknowledge the support by Institute for Basic Science (IBS-R031-D1). A.M.M. is supported by the  I+D+i project PID2020-114687GB-I00 funded by MCIN/AEI/10.13039/501100011033, by the grant Group FQM-160 and by project P20\_01247, funded by the Consejer\'{\i}a de Econom\'{\i}a, Conocimiento, Empresas y Universidad, Junta de Andaluc\'{\i}a (Spain) and by ``ERDF A way of making Europe".

\end{document}